\documentclass[prl,twocolumn,preprintnumbers]{revtex4}
\usepackage{indentfirst}
\usepackage{graphicx}
\usepackage{color}

\begin{document}

\title{Proposal for Efficient Generation of Spin-Polarized Current in Silicon}
\author{L. K. Castelano}
\email{lcastelano@physics.ucsd.edu}
\author{L. J. Sham}
\affiliation{Department of Physics, University of California-San
Diego, La Jolla, California 92093-0319}
\begin{abstract}
We propose a spin-dependent resonant tunneling structure
to efficiently inject spin-polarized current into silicon (Si).
By means of a heavily doped polycrystalline Si (Poly-Si) between the ferromagnetic metal (FM) and Si
to reduce the Schottky barrier resistance,
we estimated raising the tunneling
current density up to $10^8$Am$^{-2}$.
The small Fermi sea of the charge carriers in Si focuses the tunneling electrons to the resonant spin states within a small region of transverse momentum in the ferromagnet which creates the spin polarization of the current. Because of the large exchange splitting between the spin up and down bands, the decay of the spin current is explained in terms of scattering out of the focused beam.
The spin polarization in the current survives only if the thickness of the FM-layer is smaller than the spin-diffusion length estimated from that cause.

{\center(\today)}
\end{abstract}
 \DeclareGraphicsExtensions{.jpg, .pdf, .mps, .png, .tiff}
 \maketitle

Silicon-based devices are responsible for the processing of high volume information
 and for communication devices, thereby playing a fundamental role in existing electronics.
  Information storage, on the other hand, is in separate magnetic media. The idea of combining both functions
  in the same device
  spurred
  vast interest in spintronics with semiconductors.
  Many advantages of such devices were pointed out in recent years, \emph{e.g.}, memory nonvolatility,
  increased data processing speed, decreased power consumption, and increased integration
  densities \cite{awschalom2001}.
 The first step to implement spintronics in semiconductors consists in the ability to inject and to
  detect spin-polarized carriers in such materials. Such a task can be achieved
by making a contact between a semiconductor and a FM,
since the electrical conductivity for majority
spin and minority spin is different in the FM.
However, this type of contact produces a very low efficiency spin injection.
The difficulty is due to the difference between
the conductivities of these materials \cite{schmidt} and to circumvent such a problem, a barrier must be
included between the two media \cite{rashba}. Thus, either an insulator or a Schottky
barrier can be used to raise the efficiency of the spin injection into semiconductors;
on the other hand, the current is drastically reduced by the barriers.
The problem of spin injection efficiency into semiconductors is well understood and many
experimental results have shown its feasibility \cite{ohno,jonker,appelbaum}.
The current challenge is to create a significant electrical output signal that captures the spin information.
 In other words, to build
 an
 efficient silicon-based spintronic device, we need to find a system
  where high current and spin-polarization coexist.

In this letter, we propose a structure composed of a FM
between two heavily doped semiconductor to achieve both
spin-polarization and high total current. The operation of such a device is in a sense analogous
to the operation of the resonant tunneling diode (RTD) \cite{tsu}, therefore we will call it spin-RTD hereafter.
\begin{figure}[b!]
\centering
\includegraphics[angle=0,scale=.85]{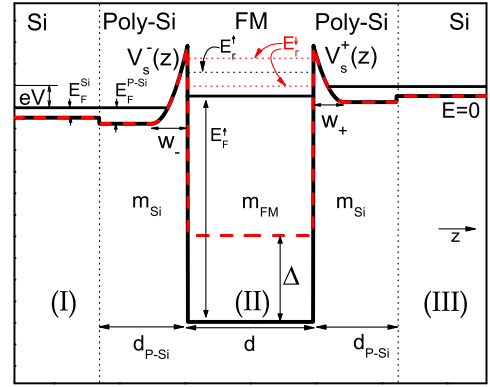}
\caption{(Color online) Schematic conduction band diagram for spin-up electrons (solid curve) and
spin-down electrons (dashed curve)
at bias $eV$.
 The Fermi energy of Poly-Si (Si) is represented by
$E_F^{P-Si}$ ($E_F^{Si}$).  d (d$_{P-Si}$) is the thickness of the
FM (Poly-Si) layer. $w_+$ and $V_s^+(z)$ ($w_{-}$ and $V_s^{-}(z)$)
designate the width and potential of the Schottky barrier in the
right (left) side, respectively. The effective mass of the FM (Si)
is given by m$_{FM}$ (m$_{Si}$). Also, resonant state with energy
$E^\uparrow_r$ ($E^\downarrow_r$) for spin-up (spin-down) electrons
is schematized. $\Delta$ is the exchange
energy, which also gives the difference between the bottom of the
two spin conduction bands of the FM.}
\end{figure}
Current in a RTD
flows only when the applied voltage reaches specific values that
correspond to resonant states within the barriers. If these resonant
states are spin-dependent, each spin component of the current will
behave differently as a function of the applied bias.
For instance, spin-down (spin-up) electrons will flow when the
applied voltage reaches the specific resonant energy $E_r^\downarrow$
($E_r^\uparrow$) [see Fig. 1], while electrons with opposite spin
will be filtered. The low density electrons in the conduction valleys
of Si serve to focus the transport electrons in the FM
region to a small part of the Fermi surface \cite{200pearson}. These
regions as filamentary channels in $k$ space together with the
ferromagnet subbands enable the relevant resonance tunneling phenomenon.
 In this work we adopt the effective mass approximation
for both metal and semiconductor. The crystal symmetry effects found important
in epitaxially grown magnetic tunnel junctions \cite{butler:mgo,mavro08} are averaged out
by the disorder in strongly doped Poly-Si. The I-V characteristics in Schottky barriers
 appear in general qualitatively accounted for by the effective mass theory \cite{duke:book}.
The band-structure of the FM is approximated by the
Stoner model, with a spin-up band and a spin-down band split by a
constant exchange $\Delta$. In Fig.~1, we present a schematic band diagram for the
proposed spin-RTD. The two thin Schottky barriers are formed by the contact between the FM
and Poly-Si. Such a material has been used
industrially as the conducting gate for MOSFET \cite{mosfet}, CMOS
\cite{cmos} and in thin-film transistors \cite{tft} applications.
Because Poly-Si can reach very high densities of dopants, we shall make use of it as an
intermediate material between the FM and Si. Also, this
material forms an ohmic contact with Si, which is located at
the edges of our spin-RTD. In the middle of Fig.~1, we represent the
two spin bands of the FM, shifted by the exchange
energy $\Delta$. Experimentally, we believe that the techniques
developed for
magnetic nanopillars \cite{nanopillars} might be very useful to
build the proposed device.

The current density for each spin component at zero temperature is determined by \cite{duke:book}
\begin{equation}\label{currt0}
    j_\sigma=\frac{em_{Si,\parallel}}{\hbar^3}\int_{E_{min}}^{E^{Si}_F} dE
\int_0^E\frac{dE_\parallel}{(2\pi)^2}D_\sigma(eV;E-E_\parallel),
\end{equation}
where $\sigma=(\uparrow,\downarrow)$, $E_{min}=(E^{Si}_F-|eV|/2)\Theta(E^{Si}_F-|eV|/2)$, $E_\parallel=\hbar^2k_\parallel^2/2m_{Si,\parallel}$, $m_{Si,\parallel}$ is the effective mass of Si parallel
to the transport direction, and
$D_\sigma(eV;E-E_\parallel)$ denotes the transmission
probability
 of the electron of spin $\sigma$
through the spin-RTD. $\Theta(x)$ is the step function.

To treat the spin relaxation of the current in the paramagnet regions of Si and Poly-Si, we use the usual spin-flip term in the spin diffusion theory for paramagnetic metals  \cite{vanson,Johnson:1993}. Although the spin flip term is also used in the multilayers of para- and ferromagnets \cite{Valet:1993}, the large exchange splitting in the ferromagnet means that $s-d$ electron scattering cannot satisfy energy conservation without concomitant change of the spatial states. Consider the common case of Si interface in the (001) direction with two pockets of conduction electrons. Their low density limits the tunneling current in the FM region to be less than 0.1~\% of the FM Fermi surface cross-section normal to (001), increased to at most 1~\% in the Poly-Si regions. We suggest that, for the electron subbands in  the quantum well of the FM, the confinement of the transport electrons near the Fermi level to small transverse momenta causes the current to be spin-filtered in resonance with a subband edge of a particular spin.   The case of two different spin paths is analogous to the case of strong spin-orbit split bands. \cite{163Vinattieri}.  The polarization  decay  in the tunneling current  is due to scattering with the  electrons  outside the tunneling current.

To obtain a rough estimate of the spin currents, we use, in FM, the exchange splitting of the two bands for the difference in the real parts of $V_\sigma(z)$ and the constant imaginary parts of $W_\sigma$ for the current decays in the $d$ tunneling electrons in the different spin channels.
Then the spin-dependent transport relaxation time is k-independent, $\tau_\sigma=\hbar/2W_\sigma$.
We argue that the inhibiting effect of disparate energy levels of the exchange-split bands in ferromagnetic metals is present at the interface, even without the semiconductor focussing effect and thus interpret
the bulk and interface measurement results in ferromagnetic metals which are generally presented as the spin-diffusion length \cite{bass}. Without the additional data for the spin-independent component of the transport time of the electron in the tunneling channel, we cannot unentangle the conductivity and the diffusion coefficient for each spin channel, as was done in the semiconductor case with optical excitation \cite{163Vinattieri}. We simply approximate the two independent exponential decaying spin-components in the accumulation layer by the same measured \cite{bass} ``spin-diffusion length''  $\ell^\sigma_{sd}$. Thus, $\tau_\sigma=\ell^\sigma_{sd}/v^\sigma_F $, where $v^\sigma_F=\sqrt{2E_F^\sigma/m_{FM}}$ is Fermi velocity in the ferromagnet.

\begin{figure}[b!]
\centering
\includegraphics[angle=0,scale=.8]{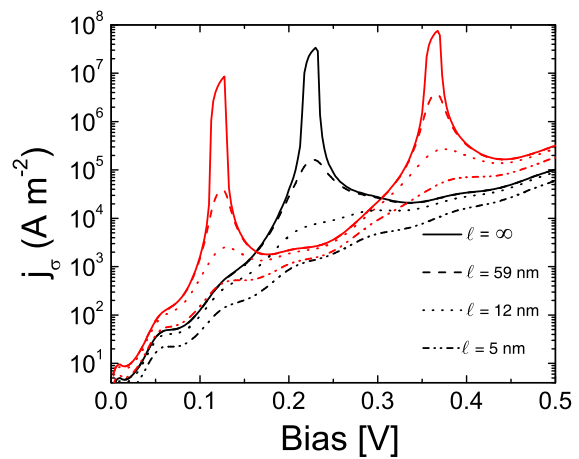} %
\caption{(Color online) The black (red or light gray) curve
represents the current density for spin-up (spin-down) electrons as
a function of applied bias. Here, we consider a fixed thickness for the FM-layer
$d=8.3$ \emph{nm} and T=0 K. The results considering different ``spin-diffusion length''
are also indicated.}
\end{figure}

The shape of the Schottky barrier
in the Poly-Si semiconductor region ($s$) is calculated in
the depletion layer approximation,
with a constant donor concentration, $N^{\text{P-Si}}_d$, in $-(w_-+a)<z<-a$ and $a<z<w_++a$,
where $a=d/2$ and the origin $z=0$ is set in the middle of the FM.
The electrostatic potential of the barriers is,
\begin{eqnarray}
V^{\pm}_{\text{s}}(z)  &=& \frac{2\pi e^2N^{\text{P-Si}}_d}{\epsilon_{\text{s}}} \left(z^2\mp
2z(a+w_{\pm})+2aw_{\pm}+a^2\right)  \nonumber \\
&+&   E_F^{\text{Si}}+V_{\text{s}}-eV/2,  \label{barriers}
\end{eqnarray}
where $\epsilon_{\text{s}}$ is the dielectric constant of the semiconductor
and plus (minus) signal refers to the right (left) side. The height of the Schottky barriers is
$V_{\text{s}}$  and the widths are $w_{\pm}=\left[\frac{\epsilon_{\text{s}}}{2\pi e^2N^{P-Si}_d}(E_F^{\text{P-Si}}+V_{\text{s}} \mp eV/2)\right]^{1/2}$.

 In our numerical calculation, we use the following
parameters: $v_F^{\uparrow}=1.26\times 10^8$ cm/s,
$v_F^{\downarrow}=4.8\times 10^7$ cm/s (corresponding to the two
exchange split bands $E_F^{\uparrow}=4.52$ eV,
$E_F^{\downarrow}=E_F^{\uparrow}-\Delta=0.66$ eV), and the effective
mass of electrons inside the FM equal to the free-electron mass
(m$_{\text{FM}}$=m$_0$) \cite{slonczewski}. For silicon, we adopt
m$_{Si,z}=0.91$ m$_0$ and m$_{Si,\parallel}=0.19$ m$_0$. For
simplicity, we assume the same effective mass of silicon for
Poly-Si. The carrier concentrations are
$N^{\text{P-Si}}_d=10^{20}$ cm$^{-3}$ and
$N^{\text{Si}}_d=5\times10^{18}$ cm$^{-3}$ for heavily doped
Poly-Si and Si, respectively. Also, we consider a Schottky barrier
height $V_s=0.65$ eV, the Fermi energy of Poly-Si and Si equal to
$E_F^{\text{P-Si}}\approx80$ meV \cite{young}, and $E_F^{\text{Si}}\approx10$ meV,
respectively. The Schottky barrier width for zero bias is $w_-=w_+\approx3.2$ \emph{nm}
and the thickness of the Poly-Si layer is fixed at $d_{\text{P-Si}}=5$ \emph{nm}.

The current density for spin-up (spin-down) electrons
calculated by Eq.~(2) is shown by the black (red or light gray) solid curve in
Fig.~2 for a fixed thickness of the FM-layer $d=8.3$ \emph{nm}.
The spin-up current density shows a maximum
(2.4$\times$10$^7$Am$^{-2}$) for an applied bias of 0.23 Volts. A
similar phenomenon is observed for the spin-down current
density, although two peaks are observed in this case.
Such behavior is the signature of the RTD devices, where the
current flows only
after
the applied bias tunes the resonant states.
\begin{figure}[b]
\centering
\includegraphics[angle=0,scale=.81]{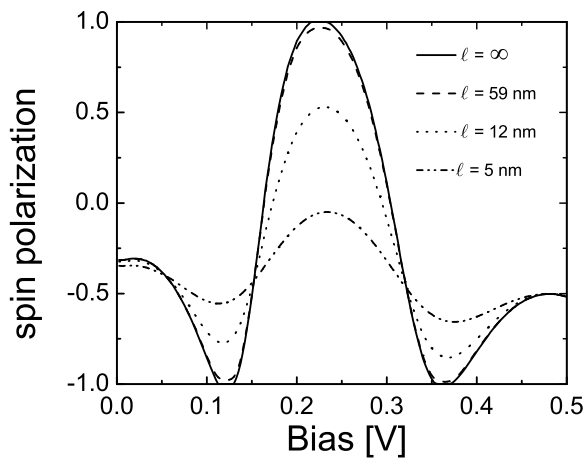} %
\caption{ Spin-polarization as a function of
applied bias. The thickness for the FM-layer
and the temperature are the same adopted earlier ($d=8.3$ \emph{nm} and T=0 K). The results considering different ``spin-diffusion length''
are also indicated.}
\end{figure}
Fig.~2 also shows the effects of partial relaxation on spin components of the
current density for different values of
the
``spin-current decay length''
corresponding to three different FM, Co
($\ell^{\text{Co}}_{\text{sd}}=$59~nm), CoFe ($\ell^{\text{CoFe}}_{\text{sd}}=$12~nm), and Fe
($\ell^{\text{Fe}}_{\text{sd}}=$5~nm) \cite{bass}. As expected, we see that
decreasing the spin current decay length reduces and broadens the current density peaks.
When the spin-decay length
is smaller than the thickness of the
FM-layer ($\ell^{\text{Fe}}_{\text{sd}}=$5~nm) we observe a large suppression in the current density peaks. The changes in the spin-dependent density current as a function of the
applied bias can be observed by the spin-polarization, defined as
$p=(j_\uparrow-j_\downarrow)/(j_\uparrow+j_\downarrow)$. Fig.~3
shows the spin-polarization as a function of the bias for a fixed
thickness for the FM-layer ($d=8.3$~nm) and the same spin-current decay lengths.
In this case of $\ell_{\text{sd}}>>d$, i.e., sufficient preservation of the spin current, the current polarization oscillates and can reach full spin polarization.  Thus, a highly spin-polarized current with either spin direction may be obtained by appropriate choice of the applied bias  on the spin-RTD.


In conclusion, we show the possibility of injection of highly spin-polarized and strong current density into silicon by employing a spin-RTD.
The focusing effect of the tunnel current by the semiconductor enables resonant tunneling dominated by a single spin subband of the FM quantum well at a given bias voltage. The possibility of polarization switching by electrical bias control may be of importance to spin devices. The spin decay through the ferromagnet is avoided by short layer width.
The low polarization generation due to the resistance mismatch between the
metal electrode and the semiconductor is mitigated by the intervening heavily doped electrodes (Poly-Si).
Finally, the fabrication of such spin-RTD is within the capabilities of current nanomagnet plus semiconductor stack fabrication.

This work is supported by U.S. Army Research Office MURI W911NF-08-2-0032 and by
CNPq (Brazil).

\end{document}